\documentclass[preprint,showpacs,preprintnumbers,amsmath,amssymb]{revtex4}

\usepackage{graphicx}% Include figure files
\usepackage{dcolumn}% Align table columns on decimal point
\usepackage{bm}% bold math

\begin{document}

%\preprint{APS/123-QED}

\title{Comparison of $H_{c2}(T,\theta)$
in Mg$_{1-x}$Al$_x$B$_2$ single crystals with the dirty-limit
two-gap theory}

\author{Heon-Jung Kim$^1$, Hyun-Sook Lee$^1$, Byeongwon Kang$^{1,\dag}$, Woon-Ha Yim$^1$, Younghun Jo$^2$, Myung-Hwa Jung$^2$, and Sung-Ik Lee$^{1,2}$}
\affiliation{$^1$National Creative Research Initiative Center for
Superconductivity and Department of Physics, Pohang University of
Science and Technology, Pohang 790-784, Republic of Korea }
%\author{Younghun Jo and Myung-Hwa Jung}
\affiliation{$^2$Quantum Material$¡¯$s Research Laboratory, Korea
Basic Science Institute, Daejeon 305-333, Republic of Korea}

\date{\today}

\begin{abstract}
{We studied the temperature and the angular dependences of the
upper critical field ($H_{c2}(T,\theta$)) of Mg$_{1-x}$Al$_x$B$_2$
single crystals ($x = 0.12$ and 0.21) and compared with the
dirty-limit two-gap theory. We found that $H_{c2}(T,\theta)$'s
were well described in a unified way by this theory. The obtained
values of the parameters indicated that as the Al concentration
was increased, anisotropic impurity scattering increased, making
the $\sigma$ bands less anisotropic. Accordingly, the temperature
dependence of the anisotropy ratio of $H_{c2}$ ($\gamma_H$)
systematically decreased, and for $x=0.21$, $\gamma_H$ was nearly
constant. Our results imply that Mg$_{1-x}$Al$_x$B$_2$ single
crystals are in dirty-limit and that two-gap nature survives until
$x = 0.21$. }
\end{abstract}

 \maketitle
%\section{Introduction}
It is now well established that MgB$_2$ is a two-gap
superconductor with two distinct energy gaps: a large gap
originating from two-dimensional $\sigma$ bands and a small gap
originating from three-dimensional $\pi$ bands
\cite{Liu01,Choi02,Souma03}. One of the main consequences of the
two-gap nature is the strong temperature dependence of the
$H_{c2}(T)$ anisotropy, $\gamma_H \equiv H_{c2}^{ab} /\ H_{c2}^c$
\cite{Lyard02}, which is not expected based on the single-gap
Ginzburg-Landau theory. Theoretical calculations show that the
strong temperature dependence of $\gamma_H$ arises from the fact
that the anisotropic $\sigma$ bands dominate $\gamma_H$ at low
temperatures while the $\pi$ bands gradually become important at
temperatures near $T_c$ \cite{Dahm03,Gurevich03,Golubov03}. The
above anomalous behavior of $\gamma(T)$ for MgB$_2$ single
crystals was confirmed by using magnetization measurements
\cite{Lyard04,HJKim04}.

When impurity scattering is increased, the above-mentioned
behaviors of $H_{c2}$ are modified. Gurevich \cite{Gurevich03},
and Golubov and Koshelev \cite{Golubov03} formulated the
dirty-limit two-gap theory for $H_{c2}$ by using the
quasiclassical Usadel equations. According to this theory, the
shape of the $H_{c2}(T)$ curve essentially depends on the
diffusivities of the $\sigma$ and the $\pi$ bands. For $T \approx
T_c$, $H_{c2}(T)$ is determined by a maximum diffusivity (cleaner
bands) between $D_{\sigma}$ and $D_{\pi}$ while $H_{c2}(0)$ is
controlled by a minimum diffusivity (dirtier bands). When the
$\sigma$ bands are dirtier, an upward curvature should appear near
$T_c$, and $\gamma_H$ should decrease with temperature. In
contrast, when the $\pi$ bands are dirtier, a huge increase in
$H_{c2}(T)$ should appear at low temperatures without an upward
curvature near $T_c$, and $\gamma_H$ should increase with
temperature.

Impurity scattering also changes $H_{c2}(\theta)$.
$H_{c2}(\theta)$ was predicted to deviate from the angular
dependence of the anisotropic one-gap Ginzburg-Landau (GL) theory,
especially near the middle-angle region.  This deviation should be
most pronounced at $T/T_c \approx 0.95$ when the parameters
supplied by band-structure calculations are used
\cite{Golubov03,Rydh04}. Even though these predictions were
quantitatively compared with $H_{c2}(\theta)$ for MgB$_2$ single
crystals and reasonable consistency was observed \cite{Rydh04},
the problem of whether the dirty-limit theory could be applied to
clean MgB$_2$ single crystals still remained. In this sense, the
dirty-limit theory has not yet been verified unambiguously for
single crystals in the dirty limit, especially for the
orientational dependence of $H_{c2}$.

In this paper, we report the effect of Al doping, as deduced from
resistance measurements at various angles $\theta$ between $H$ and
the $c$-axis, on $H_{c2}(T,\theta)$ of Al-doped MgB$_2$ single
crystals. This directional study of the resistance was possible
due to success in growing flat and regular-shaped  Mg$_{1-
x}$Al$_x$B$_2$ single crystals with values of $x$ up to 0.21 and
with $T_c$ = 25.5 K. We found that two-gap superconductivity in
MgB$_2$ was drastically affected by the Al doping and that key
features predicted by the dirty-limit two-gap theory were
observed. Our main observations are the following: (1) As the Al
concentration increases, the residual resistivity ($\rho_0$)
greatly increases, implying that Al substitution enhances impurity
scattering and that the Al-doped samples are in the dirty-region.
(2) $H_{c2}(T)$ can be consistently explained within the
dirty-limit two-gap theory up to $x = 0.21$, even though
$H_{c2}(0)$ decreases with Al concentration. (3) The $\gamma_H(T)$
systematically decreases and for $x=0.21$, $\gamma_H$ is virtually
temperature-independent. (4) The $H_{c2}(\theta)$ for $x$ = 0.12
showed a clear deviation from the behavior predicted by the
anisotropic GL theory, which is a strong indication of the two-gap
nature in MgB$_2$. However, for $x$ = 0.21, this deviation became
very small. The values of the obtained parameters suggest that
impurity scattering is enhanced in the $\pi$ bands, especially
along the $c$ direction and that the anisotropy of the $\sigma$
bands is significantly reduced.

%Experimentally, the effect of impurity scattering due to Al
%substitution has been controversial. Regarding the effect of Al
%substitution on $H_{c2}(T)$, Putti {\it et al.} \cite{Putti05} and
%Angst {\it et al.} \cite{Angst05} insisted that $H_{c2}(T)$ could
%be explained in the clean limit by a shift in the Fermi level due
%to electron doping while Park {\it et al.} \cite{MSPark05} showed
%that the dirty-limit two-gap theory could explain their
%$H_{c2}(T)$ data. However, all these studies have limitations in
%that detailed angle-dependent investigations, which are possible
%only when well-shaped MgB$_2$ single crystals are used, were not
%performed.

%\section{Experiments}
Mg$_{1-x}$Al$_x$B$_2$ single crystals with $x$ = 0.12 and 0.21
were grown under high-pressure conditions \cite{Kijoon02,BWKang04}
and were characterized and patterned as in
\cite{Kijoon02,BWKang04}. Two sets of samples with clean, shiny
surface were investigated for each Al concentration.
%
%The Al concentrations $x$, determined by using both electron-probe
%X-ray microanalysis and electron dispersive X-ray spectroscopy,
%were 12 $\pm$ 1\% and 21 $\pm$ 2\%, respectively. Low field
%magnetization measurements were performed using a superconducting
%quantum interference device magnetometer (Quantum Design, MPMSXL).
For the resistance measurements, well-shaped single crystals with
both sides flat were selected from numerous samples. The
temperature and the angular dependences of the resistance were
measured from 0 to 9 T by using the AC transport option in a PPMS
Quantum Design system.

%\section{Results and discussion}
Figure~\ref{fig1} shows the resistivity $\rho$ of the
Mg$_{1-x}$Al$_x$B$_2$ single crystals ($x$ = 0, 0.12, and 0.21) as
a function of temperature. As the Al concentration increases,
$T_c$ decreases. The $T_c$'s are 30.8 K and 25.5 K for $x$ = 0.12
and $x$ = 0.21, respectively. The data for $x = 0$ were taken from
Ref. \cite{Lyard02} and $T_c$ of this sample was around 37 K.
Previously, for MgB$_2$ single crystals, the resistance was
reported to follow the Bloch-Gr\"{u}neisen (BG) formula with a
Debye temperature of $\Theta_D \sim 1100$ K \cite{Kijoon02}. This
implied that the normal-state transport properties were well
described by an electron-phonon interaction without considering an
electron-electron interaction. To check whether this is the case
in Al-doped single crystals, we fitted the $\rho(T)$ data with the
BG formula, where fitting paramters are $\Theta_D$ and residual
resistivity $\rho_0$. The solid lines in the figure are the BG
theoretical curves and describes the $\rho(T)$ data well. The
value of $\Theta_D$ in Al-doped single crystals is $\sim$ 1000 K,
which is similar to that of MgB$_2$ single crystals. $\rho_0$
increases monotonically with doping, and the fitted values of
$\rho_0$ are 1.63, 21.4, and 32.2 $\mu \Omega$ cm for $x$ = 0.0,
0.12, and 0.21, respectively. The inset of Fig. \ref{fig1} shows
the normalized low-field magnetization for zero-field-cooled state
of Mg$_{1-x}$Al$_x$B$_2$ single crystals extracted from the same
batch of single crystals as was used for the resistivity
measurements. The $T_c$'s determined from the resistivity and from
the low-field magnetization were virtually the same.

Figure \ref{fig2} (a) and (b) show, as an example, the temperature
dependences of the resistances of the $x$ = 0.12 sample for $H
\parallel c$ and $ \parallel ab$, respectively. As with
MgB$_2$ single crystals, this sample shows surface
superconductivity: As the temperature decreases, the resistance
first decreases linearly and then suddenly drops to zero. In the
region of linear decrease, the resistance depends on the applied
current, and a higher current induces a higher resistance. The
drop in the resistance indicates the onset of bulk
superconductivity. Particularly at high currents ($I$ = 3 mA) for
$H \parallel c$, a peak, which is absent at low currents ($I$ = 1
mA), appears. The current dependence of this peak suggests that it
is due to the peak effect, observed in MgB$_2$ single crystals
\cite{Welp03}. The upper critical fields can be determined
unambiguously as the points where the resistance drops to zero in
the curves for I = 1mA. Those points are indicated by the arrows.

In Fig. \ref{fig3}(a), $H_{c2}^{c}(T)$ and $H_{c2}^{ab}(T)$ for
$x$ = 0.12 and 0.21 are plotted, where $H_{c2}^{c}(T)$ and
$H_{c2}^{ab}(T)$ are $H_{c2}(T)$'s for $H \parallel c$ and for $H
\parallel ab$, respectively. For comparison, we also insert $H_{c2}(T)$ for $x$ = 0.0,
which was taken from Ref. \cite{Lyard02}. Interestingly, both
$H_{c2}^{c}(T)$ and $H_{c2}^{ab}(T)$ decrease with increasing Al
doping. As a result, the extrapolated $H_{c2}^{c}(0)$ and
$H_{c2}^{ab}(0)$ are reduced. While the decrease in
$H_{c2}^{ab}(0)$ is consistent with the results for
polycrystalline samples, the decrease in $H_{c2}^{c}(0)$ is not.
In a study by Angst {\it et al.}, a small increase in
$H_{c2}^{c}(0)$ was observed at an Al doping of 10 \%
\cite{Angst05}. By comparing $H_{c2}(0)$ in both Al- and C-doped
MgB$_2$, they concluded that in Al-doped samples, the shift in the
Fermi level was dominant in determining $H_{c2}(T)$ while in
C-doped samples, disorder played a major role. However, in light
of the huge increase in $\rho_0$, the effects of disorder are not
negligible and should be taken into account. Another clue to the
degree of dirtiness in Mg$_{1-x}$Al$_x$B$_2$ single crystals for
$x$ = 0.12 and $x$ = 0.21 can be found in the shape of
$H_{c2}^{c}(T)$ near $T_c$. While MgB$_2$ single crystals show a
linear decrease in $H_{c2}^{c}(T)$ near $T_c$, close inspection
reveals that an upward curvature gradually appears with Al doping.
This becomes even clearer if $H_{c2}^c(T)$ for $x = 0$ is compared
with that for $x=0.21$. The upward curvature is consistent with
the two-gap dirty-limit theory.

Since the variations in $H_{c2}(T)$ with Al doping appear to agree
well with the two-gap theory, we quantitatively analyzed our
$H_{c2}(T)$ data by using the dirty-limit theory
\cite{Gurevich03,Golubov03}. For $x$ = 0, the dirty limit model
may be inappropriate because pure MgB$_2$ crystals are considered
to be in the clean limit \cite{Yelland02}. If interband impurity
scattering is assumed to be zero, $H_{c2}(T)$ for $H \parallel c$
is given by
\begin{eqnarray}
& a_0\left[ \text{ln } t + U(h)\right] \left[\text{ln }t+U(\eta
h)\right]
+ a_2 \left[\text{ln }t+U(\eta h)\right] & \nonumber \\
& + a_1 \left[\text{ln } t + U(h)\right]=0, & \label{eq1}
\end{eqnarray}
where $t = T/T_c$, $U(x)=\Psi(1/2+x)-\Psi(x)$, $\Psi(x)$ is the
Euler digamma function, $h= H_{c2} D_{\sigma}^{ab}/2\phi_0 T$,
$\phi_0$ is the magnetic flux quantum, $\eta =
D_{\pi}^{ab}/D_{\sigma}^{ab}$, $D_{\sigma,\pi}^{ab}$ is the
in-plane electron diffusivity of the $\sigma$ and the $\pi$ bands,
 and $a_{0,1,2}$ are constants derived from the electron-phonon coupling
 constants ($\lambda_{mn}^{ep}$) and the Coulomb pseudopotentials
 ($\mu_{mn}$). The precise definitions of $a_{0,1,2}$ can be found in Ref.
 6. For $H \parallel ab$, the in-plane diffusivities in Eq. \ref{eq1} can be replaced
 by $[D_{\sigma,\pi}^{ab} D_{\sigma,\pi}^{c}]^{1/2}$, where $D_{\sigma,\pi}^c$ are the out-of-plane electron
diffusivities of the $\sigma$ and the $\pi$ bands, respectively.
Equation \ref{eq1} can be generalized to the anisotropic case of
an inclined field by replacing the diffusivities with the
angle-dependent diffusivities $D_{\sigma}(\theta)$ and
$D_{\pi}(\theta)$ for both bands, where
$D_{\sigma,\pi}(\theta)=[(D_{\sigma,\pi}^{ab})^2 \cos^2 \theta +
D_{\sigma,\pi}^{ab} D_{\sigma,\pi}^c \sin^2\theta]^{1/2}$.
 For the four input parameters
$\lambda_{mn} = \lambda_{mn}^{ep} - \mu_{mn}$ at each Al doping
level, which reflects the change in the electronic structure by
electron doping, we used the values determined from
first-principle calculations \cite{Umm04}, and we obtained the
numerical value of the diffusivity for each band.

In our samples, the interband impurity scattering is believed not
to be significant or, if any, to be negligible to the first
approximation. This is because the interband impurity scattering
was predicted to eliminate the distinction of each superconducting
gap, destructing two-gap features \cite{Golubov97}. Therefore, the
upward curvature, which is the hallmark of the two-gap
superconductivity would have not been observed, if interband
impurity scattering were significant.

The solid lines in Fig. \ref{fig3}(a) present the theoretical
two-gap dirty-limit curves of $H_{c2}(T)$ for $x$ = 0.12 and 0.21.
The optimized values of $D_{\sigma}^{ab,c}$, $D_{\pi}^{ab,c}$, and
 $H_{c2}^{ab,c}(0)$ from the fits are summarized in Table
\ref{tab:table1}. The upward curvature observed near $T_c$ for $x$
= 0.12 and 0.21, which is typical when $\sigma$ bands are dirtier
than $\pi$ bands \cite{Gurevich03}, may indicate dirtier $\sigma$
bands. If the $\pi$ bands are dirtier than the $\sigma$ bands, the
upward curvature near $T_c$ should disappear; instead, a huge
increase in $H_{c2}(T)$ should appear at low temperatures. The
dashed line for $x = 0$ is a guide to eyes.

Quantitatively, the values of $D_{\sigma}^{ab,c}$ and
$D_{\pi}^{ab,c}$ really prove dirty $\sigma$ bands
($D_{\sigma}^{ab,c} \ll D_{\pi}^{ab,c}$), which is consistent with
the shape of $H_{c2}(T)$. Dirty $\sigma$ bands were also observed
in Al-doped MgB$_2$ polycrystalline samples \cite{MSPark05}. The
electron diffusivity along the $c$ direction in the $\pi$ bands is
noted to decrease with Al doping while that in the $ab$ plane
virtually does not change. This originates from pronounced
impurity scattering in the $\pi$ bands as the Al concentration is
increased. The pronounced impurity scattering, however, is not
isotropic as is normally assumed. Along the $c$ direction,
impurity scattering is more enhanced than in the $ab$ plane.
Similarly, Al doping influences impurity scattering in the
$\sigma$ bands. In this case, the electron diffusivity along the
$c$ direction increases with Al doping while that in $ab$ plane
virtually is unchanged. Consequently, the $\sigma$ bands become
more isotropic, which is reflected in the ratio
$D_{\sigma}^{ab}/D_{\pi}^c$ and this value decreases as Al content
increases. The isotropization of the $\sigma$ bands is believed to
be due to not only the anisotropic impurity scattering but also
the change in the electronic structure that Al doping induces.

The same set of electron diffusivities as in Table
\ref{tab:table1} can explain $H_{c2}(\theta)$ for $x$ = 0.12 and
0.21, as shown in Fig. \ref{fig3}(b). The solid lines indicate the
theoretical curves calculated from the dirty-limit two-gap theory.
The dotted lines are the theoretical curves of the one-gap GL
model. The error bars in this data are comparable to or less than
the symbol size. The two-gap theory describes the data better than
the GL model for $x = 0.12$.  For $x$ = 0.12, a small difference
between the two-gap theory and the GL model is apparent and as
predicted, is most pronounced at the middle-angle regions. This is
a strong indication of the two-gap nature of Al-doped MgB$_2$
single crystals. This behavior is very similar to that of MgB$_2$
single crystals, where a deviation from GL behavior was observed
to be peaked at $T \approx 0.8 T_c$. For $x$ = 0.21, the
difference between the two-gap theory and the anisotropic GL model
is very tiny, as is the case for the temperatures we investigated.
Despite the indistinction between the anisotropic GL model and the
two-gap theory for this doping, the shapes of the $H_{c2}(T)$
curves and the values of the fitted diffusivities guarantee the
existence of two distinct gaps. If the sample for $x = 0.21$
followed the one-gap GL model, the upward curvature would not be
observed.

Finally, $\gamma_H(T)$ for $x$ = 0, 0.12, and 0.21, extracted from
the $H_{c2}(T,\theta)$ data, are plotted as functions of the
reduced temperature $T/T_c$ in the inset of Fig. \ref{fig3}(a).
The values of $\gamma_H$ are systematically reduced, and for $x$ =
0.21, $\gamma_H$ is virtually temperature-independent at high
temperatures, slightly increasing at low temperatures. The
$\gamma_H$ at low temperatures significantly changes with Al
doping and the $\gamma_H$'s merge to 2 $-$ 2.5 at $T = T_c$ for
all doping levels. This behavior is thought to result from the
isotropization of the $\sigma$ bands. The decreasing tendency of
$\gamma_H$ with increasing temperature for $x$ = 0.12 and 0.21 is
in good agreement with the case of dirty $\sigma$ bands, predicted
by using the dirty-limit two-gap theory.

If the effects of impurity scattering can be ignored in $x$ = 0.12
and 0.21 single crystals, $H_{c2}$ will evolve according to
changes in the electronic structure and in the lattice constant.
Among these, the main effect is due to changes in the electronic
structure caused by doping with electrons, resulting in a shift of
Fermi level $E_F$ to higher energies. At moderate doping levels,
where a rigid band model is valid, an increase in $E_F$ modifies
the band-averaged Fermi velocities, primarily in the $\sigma$
bands and the $\gamma_H(0)$, which is $\gamma_{v_F} \equiv
v^{ab}_{F,\sigma}/v^{c}_{F,\sigma}$ in the clean limit. Here,
$v^{ab (c)}_{F,\sigma}$ is the in-plane (out-of-plane) Fermi
velocity of the $\sigma$ bands. According to the calculation by
Putti {\it et al.} \cite{Putti05}, $v^{c}_{F,\sigma}$ remains
approximately constant while $v^{ab}_{F,\sigma}$ substantially
decreases with Al doping for $ x < 0.3$. At doping levels of $x$ =
0.0, 0.12, and 0.21, that calculation produced $\gamma_{v_F}$ =
5.6, 5, and 4.2, respectively. The value at $x$ = 0.0 is nearly
consistent with $\gamma_H(0)$ estimated from the experimental
data, as shown in the inset of Fig. \ref{fig3}(a). In contrast,
the values at $x$ = 0.12 and 0.21 are significantly larger than
the estimated $\gamma_H(0)$. In fact, the $\gamma_H$(0)'s at $x$ =
0.12 and 0.21 are better represented by the parameter
$\gamma_{\sigma} \equiv \sqrt{D_{\sigma}^{ab}/D_{\sigma}^{c}}$,
which contains information on not only the Fermi velocity but also
impurity scattering. Therefore, as we said before,
Mg$_{1-x}$Al$_x$B$_2$ single crystals ($x$ = 0.12 and 0.21) are in
the dirty limit with anisotropic impurity scattering. This is in
sharp contrast to the conclusions for Al-doped MgB$_2$
polycrystalline samples \cite{Angst05,Putti05}. Those
polycrystalline samples might have less impurities than single
crystals, which is very improbable in normal situations. It is
noted that while $\gamma_{\sigma}$ decreases with Al doping,
$\gamma_{\pi}$ increases from 1.1 to 1.8

In fact, the electron diffusivities are related to the value of
resistivity by the relation of $1/\rho \propto N_{\sigma}
D_{\sigma}+ N_{\pi} D_{\pi}$\cite{Gurevich03}, where $N_{\sigma}$
and $N_{\pi}$ are partial densities of state in $\sigma$ and $\pi$
bands, respectively. In the present case, since the electron
diffusivities in the $\pi$ bands are larger than those in the
$\sigma$ bands, the electron diffusivities in the $\pi$ bands
determine the resistivities of our samples and resistivity should
increase with $x$. This tendency holds in our samples. We
calculated the values of resistivities by using the obtained
diffusivity values and the partial densities of state calculated
by Ummarino {\it et al.} \cite{Umm04} and obtained 10 and 12
$\mu\Omega cm$ for $x = 0.12$ and $x = 0.21$, respectively. The
discrepancy of the absolute values, especially for $x = 0.21$
might originate from a large error in calculating the resistivity
of small-sized samples.

%\section{Summary}
To summarize, we investigated the effect of Al substitution on
$H_{c2}(T,\theta)$ of MgB$_2$ single crystals. From an analysis of
$H_{c2}(T,\theta)$ within the dirty-limit two-gap theory, we found
that Al substitution influenced the electronic structure
complexly; in the $\pi$ bands, it increased impurity scattering
along the $c$ direction while it made the $\sigma$ bands less
anisotropic. Accordingly, $\gamma_H(T)$ was systematically
decreased and for $x=0.21$, $\gamma_H$ was virtually
temperature-independent. The isotropization, especially of the
$\sigma$ bands, originates not only from increased anisotropic
impurity scattering but also from electron doping. In
$H_{c2}(\theta)$, we also observed a strong indication of the
dirty-limit two-gap nature of Al-doped MgB$_2$.

\begin{acknowledgments}
This work is supported by the Ministry of Science and Technology
of Korea through the Creative Research Initiative Program and by
the Asia Pacific Center for Theoretical Physics. This work was
partially supported by the National Research Laboratory Program
through the Korea Institute of Science and Technology Evaluation
and Planning.
\end{acknowledgments}

$^\dag$present address : Department of Physics, Chungbuk National
University, Cheongju 361-763, Republic of Korea

\newpage
\begin{table}
\caption{\label{tab:table1} Al content $x$, upper critical fields
$H_{c2}^{ab(c)}(0)$, electron diffusivities along the $ab$ plane
(the $c$ axis) in the $\sigma$ and the $\pi$ bands,
$D^{ab(c)}_{\sigma}$ and $D^{ab(c)}_{\pi}$, obtained by fitting
the $H_{c2}(T)$ data to the dirty-limit model.}
\begin{ruledtabular}
\begin{tabular}{ccccccc}
$x$ &$H_{c2}^{ab}$(0) (T) & $H_{c2}^c$(0) (T) & $D_{\sigma}^{ab}$
($m^2s^{-1}$)
 & $D_{\sigma}^c$ ($m^2s^{-1}$)& $D_{\pi}^{ab}$ ($m^2s^{-1}$) & $D_{\pi}^c$ ($m^2s^{-1}$) \\
\hline
0.12   & 9.3 & 2.7 & $7.6 \times 10^{-4}$ & $5.9 \times 10^{-5}$  & $3.7 \times 10^{-3}$ & $3.0 \times 10^{-3}$ \\
0.21 & 5.6 & 2.3 & $6.0 \times 10^{-4}$ & $1.0 \times 10^{-4}$   & $4.8 \times 10^{-3}$ & $1.4 \times 10^{-3}$ \\
\end{tabular}
\end{ruledtabular}
\end{table}
\newpage
\begin{figure}
\caption{Temperature dependence of the resistivity for
Mg$_{1-x}$Al$_x$B$_2$ single crystals ($x$ = 0.0, 0.12, and 0.21).
The solid lines are theoretical curves of the BG formula. The
inset shows the normalized low-field magnetization in the
zero-field-cooled state. } \label{fig1}
\end{figure}

\begin{figure}
\caption{Temperature dependence of the resistance for (a) $H
\parallel c$ and (b) $H \parallel ab$.} \label{fig2}
\end{figure}

\begin{figure}
\caption{(a) Temperature dependence of $H_{c2}$ for
Mg$_{1-x}$Al$_x$B$_2$ single crystals ($x$ = 0.0, 0.12, and 0.21).
Open symbols represent $H_{c2}(T)$ for $H \parallel c$ and closed
symbols represent $H_{c2}(T)$ for $H \parallel ab$. The data for
$x$ = 0.0 were taken from Ref. \cite{Lyard02}. The inset shows
temperature dependence of $\gamma_H$. The open triangle is
$\gamma_{v_F} \equiv v^{ab}_{F,\sigma}/v^{c}_{F,\sigma}$, and the
open circle and squre are $\gamma_{\sigma} \equiv
\sqrt{D_{\sigma}^{ab}/D_{\sigma}^{c}}$'s.
%It can be clearly seen
%that for $x$ = 0.0, $\gamma_{v_F}$ represents $\gamma_H(0)$ while
%for $x$ = 0.12 and 0.21, $\gamma_{\sigma}$ represents
%$\gamma_H(0)$.
(b) Angular dependence of $H_{c2}$. The solid lines
are the theoretical curves for the dirty-limit two-gap model, and
the dotted lines are those for the Ginzburg-Landau theory. }
\label{fig3}
\end{figure}


\begin{thebibliography}{10s}

\bibitem{Liu01}
Amy Y. Liu, I. I. Mazin, and Jens Kortus, Phys. Rev. Lett. {\bf
87}, 087005 (2001).

\bibitem{Choi02}
Hyoung Joon Choi, David Roundy, Hong Sun, Marvin L. Cohen, Steven
G. Louie, Nature {\bf 418}, 758 (2002).

\bibitem{Souma03}
S. Souma, Y. Machida, T. Sato, T. Takahashi, H. Matsui, S.-C.
Wang, H. Ding, A. Kaminski, J. C. Campuzano, S. Sasaki, K.
Kadowaki, Nature {\bf 423}, 65 (2003).

\bibitem{Lyard02}
L. Lyard, P. Samuely, P. Szabo, T. Klein, C. Marcenat, L. Paulius,
K. H. P. Kim, C. U. Jung, H.-S. Lee, B. Kang, S. Choi, S.-I. Lee,
J. Marcus, S. Blanchard, A. G. M. Jansen, U. Welp, G. Karapetrov,
and W. K. Kwok, Phys. Rev. B {\bf 66}, 180502(R) (2002).

\bibitem{Dahm03}
T. Dahm and N. Schopohl, Phys. Rev. Lett. {\bf 91}, 017001 (2003).

\bibitem{Gurevich03}
A. Gurevich, Phys. Rev. B {\bf 67}, 184515 (2003).

\bibitem{Golubov03}
A. A. Golubov and A. E. Koshelev, Phys. Rev. B {\bf 68}, 104503
(2003).

\bibitem{Lyard04}
L. Lyard, P. Szabo, T. Klein, J. Marcus, C. Marcenat, K. H. Kim,
B. W. Kang, H. S. Lee, and S. I. Lee, Phys. Rev. Lett. {\bf 92},
057001 (2004).

\bibitem{HJKim04}
Heon-Jung Kim, Byeongwon Kang, Min-Seok Park, Kyung-Hee Kim, Hyun
Sook Lee, and Sung-Ik Lee, Phys. Rev. B {\bf 69}, 184514 (2004).

\bibitem{Rydh04}
A. Rydh, U. Welp, A. E. Koshelev, W. K. Kwok, G. W. Crabtree, R.
Brusetti, L. Lyard, T. Klein, C. Marcenat, B. Kang, K. H. Kim, K.
H. P. Kim, H.-S. Lee, and S.-I. Lee, Phys. Rev. B {\bf 70}, 132503
(2004).

\bibitem{Kijoon02}
Kijoon H. P. Kim, Jae-Hyuk Choi, C. U. Jung, P. Chowdhury,
Hyun-Sook Lee, Min-Seok Park, Heon-Jung Kim, J. Y. Kim, Zhonglian
Du, Eun-Mi Choi, Mun-Seog Kim, W. N. Kang, Sung-Ik Lee, Gun Yong
Sung, and Jeong Yong Lee, Phys. Rev. B {\bf 65}, 100510(R) (2002).

\bibitem{BWKang04}
Byeongwon Kang, Heon-Jung Kim, Min-Seok Park, Kyung-Hee Kim, and
Sung-Ik Lee, Phys. Rev. B {\bf 69}, 144514 (2004).

\bibitem{Welp03}
U. Welp, A. Rydh, G. Karapetrov, W. K. Kwok, G. W. Crabtree, Ch.
Marcenat, L. Paulius, T. Klein, J. Marcus, K. H. P. Kim, C. U.
Jung, H.-S. Lee, B. Kang, and S.-I. Lee, Phys. Rev. B {\bf 67},
012505 (2003).

\bibitem{Angst05}
M. Angst, S. L. Bud'ko, R. H. T. Wilke, and P. C. Canfield, Phys.
Rev. B {\bf 71}, 144512 (2005).

\bibitem{Yelland02}
E. A. Yelland, J. R. Cooper, A. Carrington, N. E. Hussey, P. J.
Meeson, S. Lee, A. Yamamoto, and S. Tajima, Phys. Rev. Lett. {\bf
88}, 217002 (2002).

\bibitem{Umm04}
G. A. Ummarino, R. S. Gonnelli, S. Massidda and A. Bianconi,
Physica C {\bf 407}, 121 (2004).

\bibitem{Golubov97}
A. A. Golubov and I. I. Mazin, Phys. Rev. B {\bf 55}, 15146
(1997).

\bibitem{MSPark05}
Min-Seok Park, Heon-Jung Kim, Byeongwon Kang and Sung-Ik Lee,
Supercond. Sci. Technol. {\bf 18}, 183 (2005).

\bibitem{Putti05}
M. Putti, C. Ferdeghini, M. Monni, I. Pallecchi, C. Tarantini, P.
Manfrinetti, A. Palenzona, D. Daghero, R. S. Gonnelli, and V. A.
Stepanov, Phys. Rev. B {\bf 71}, 144505 (2005).


\end{thebibliography}
\end{document}